\documentclass[global,twocolumn]{svjour}
\usepackage{graphicx}
\journalname{Applied Physics B}
\begin{document}

\title{Atom lithography without laser cooling}
\author{B.\ Smeets \inst{1,3} \and P.\ van der Straten \inst{2} \and T.\ Meijer \inst{1}
\and C.G.C.H.M.\ Fabrie \inst{1} \and K.A.H.\ van Leeuwen \inst{1}}
\institute{Department of Applied Physics, Eindhoven University of Technology,
P.O. Box 513, 5600 MB Eindhoven, The Netherlands
\email{K.A.H.v.Leeuwen@tue.nl} \and Atom Optics and Ultrafast Dynamics,
Utrecht University, P.O. Box 80000, 3508 TA Utrecht, The Netherlands
 \and present address: ASML, P.O. Box 324, 5500 AH Veldhoven, The Netherlands}

\date{\today}

\maketitle

\begin{abstract}
Using direct-write atom lithography, Fe na\-no\-lines are deposited with a
pitch of 186~nm, a full width at half maximum (FWHM) of 50~nm, and a height
of up to 6~nm. These values are achieved by relying on geometrical
collimation of the atomic beam, thus without using laser collimation
techniques. This opens the way for applying direct-write atom lithography to
a wide variety of elements.\\

\end{abstract}

\section{Introduction}
\begin{figure}[h]
\center
\includegraphics[width=50mm]{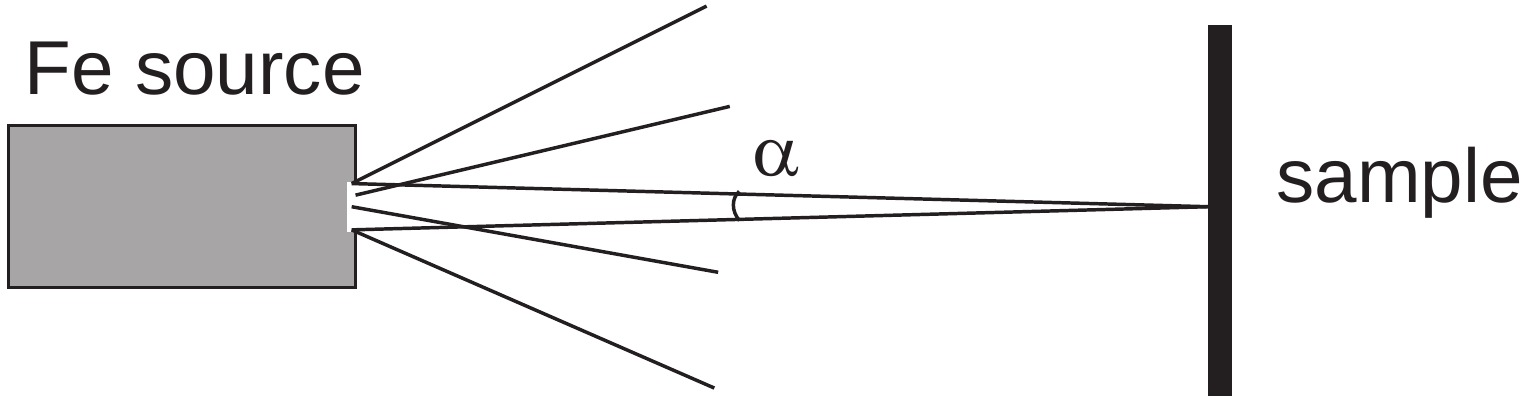}
\caption{Geometrical collimation. At each spot on the substrate, atoms arrive
with an angular spread $\alpha$, defined by the nozzle size and the distance between
source and sample.} \label{geo_coll}
\end{figure}

In direct-write atom lithography, also known as laser-focused deposition,
atoms are focused into a periodic pattern by a standing light wave. This
technique has been applied to Na \cite{art1}, Cr \cite{art2a,art2b}, Al
\cite{art3}, Yb \cite{art4}, and Fe \cite{art5,art6}. Just as for the case of
focusing light, a well collimated atomic beam is crucial for atom
lithography. An atomic beam arriving at the lens with a local angular spread
$\alpha$ (see Fig.~\ref{geo_coll}) will result in broadening of the deposited
features. To obtain a well collimated atomic beam, laser cooling techniques
are generally used. However, effective laser cooling requires a closed
optical transition. For numerous elements a closed transition is not
accessible with present lasers, or hyperfine splitting is present, and the
use of several repumping lasers becomes a necessity. For Fe, for instance, an
accessible closed transition from the ground state does not exist, resulting
in the loss of focusable atoms when applying laser cooling \cite{art8}.

It is because of these drawbacks that we decided to laser focus Fe without
the use of any laser collimation techniques, simply relying for this part on
geometrical collimation. We will show that in this way periodic grids of Fe
nanolines with a FWHM of 50~nm, a height of up to 6~nm, and a period of
186~nm can still be deposited. Thus, we prove that atoms can be focused
without using laser collimation. This makes the technique suitable for
direct-write atom lithography of any species with transitions of sufficient
strength that are not necessarily closed. Especially for atom lithography of
technologically relevant elements such as Ga \cite{art9}, In
\cite{art10,Kloeter2008}, and Si \cite{art11}, for which laser collimation
imposes large experimental difficulties, this is a step forward. Also, laser
focusing of molecules may be feasible this way.

For an earlier review of both direct-write and resist-based atom lithography,
see Ref.~\cite{Meschede2003}.

As the experiments are time-consuming, it is important to choose the
experimental parameters carefully \textit{a priori}. This is facilitated by
the availability of an efficient and realistic simulation procedure for the
atomic trajectories. Also, a comparison with simulations is essential for the
careful analysis of the experimental results. We use a fast semiclassical
Monte-Carlo procedure based on the dressed state model~\cite{art13} that
includes the effects of diffusion induced by spontaneous emission as well as
of non-adiabatic transitions between dressed states.

In Sect.\ \ref{simulations} our procedure for simulating standing-wave laser
focusing is described in more detail. In Sect.\ \ref{expsetup} the
experimental setup is described. The experimental results of atom lithography
without laser collimation are compared with the outcome of the simulations in
Sect.\ \ref{exp}, followed by the conclusions in Sect.\ \ref{concl}.

\section{Simulations}
\label{simulations}

Our simulation is based on the dressed state model introduced by
Dalibard and Cohen-Tannoudji \cite{art13}. In this model one
considers the eigenstates of the complete atom plus laser light field system.
The energies of these states are given by:
\begin{eqnarray}
E_{n,+} = (n+1)\hbar\omega_{L} -\frac{\hbar\Delta}{2} +
\frac{\hbar\Omega}{2},\nonumber\\
E_{n,-} = (n+1)\hbar\omega_{L} -\frac{\hbar\Delta}{2} -
\frac{\hbar\Omega}{2},
\end{eqnarray}
with $n$ the number of photons in the light field, $\omega_{L}$
the frequency of the light field, $\Delta=\omega_{L}-\omega_{0}$
the detuning of the light field from the atomic resonance
$\omega_{0}$. The frequency separation between the dressed states
$\Omega$ is given by:
\begin{equation}
\Omega = \sqrt{\omega_R^{2} + \Delta^{2}},
\end{equation}
with $\omega_{R}= \Gamma\sqrt{I/2I_{s}}$ the Rabi-freqency, where
$\Gamma$ represents the natural linewidth, $I$ the light
intensity, and $I_{s}$ the saturation intensity. The dressed
states are a linear superposition of the ground and excited
states:
\begin{eqnarray}\label{eq:states}
|n,+\rangle  =& \cos(\theta) |e,n\rangle &+ \sin(\theta) |g,n+1\rangle, \nonumber\\
|n,-\rangle  =& -\sin(\theta) |e,n\rangle &+ \cos(\theta) |g,n+1\rangle.
\end{eqnarray}
Here $\theta$ is defined by:
\begin{equation}\label{eq:theta}
\cos (2\theta) = -\Delta/\Omega.
\end{equation}

In our simulation, atoms move as classical point particles in a potential
field given by the dressed-state energy shift $\pm
\frac{\hbar\Omega(\vec{r})}{2}$, where the sign indicates the dressed state
the system is currently in. The light mask is described as a one-dimensional
Gaussian standing wave along the $x$-axis with a waist $w$ and maximum
on-axis intensity $I_0$:
\begin{equation}\label{eq:intensity}
I (\vec r) = I_0 \sin^{2}(kx) \exp(-2\frac{z^{2}}{w^{2}}).
\end{equation}
Here $k = 2\pi/\lambda$ is the wavenumber of the light and the propagation
direction of the atoms is assumed to be along the $z$-axis. As the atoms'
kinetic energy is much larger than the potential height of the light mask we
can neglect its effect on the longitudinal motion, so that the atom's motion
is described by a one-dimensional Newtonian equation of motion in the
$x$-direction, which is solved numerically.

Outside the standing wave the intensity $I$ of the light is zero and the
ground state is $|+\rangle$ for $\Delta>0$ and $|-\rangle$ for $\Delta<0$.
Unless the atoms undergo a spontaneous or non-adiabatic transition, they will
stay in their initial state.

The effects of spontaneous emission, and hence diffusion in the atom's
motion, are included in the model by evaluating the probability that an atom
has undergone spontaneous emission at regular time intervals. By far the
largest contribution to the diffusion is caused by emissions that change the
sign of the dressed state, causing an instant reversal of the force on the
atom. The relevant transition rates are given by:
\begin{eqnarray}
\Gamma_{+-}=\Gamma \cos^{4}\theta \\
\Gamma_{-+}=\Gamma \sin^{4}\theta,
\end{eqnarray}
with $\theta$ defined in Eq. \ref{eq:theta}. The probability of a dressed
state transition due to spontaneous emission is given by $\tau \Gamma_{+-}$
or $\tau \Gamma_{-+}$, depending on the initial dressed state, with $\tau$
the time interval.

\begin{figure}[htbp]
\center
\includegraphics[width=85mm]{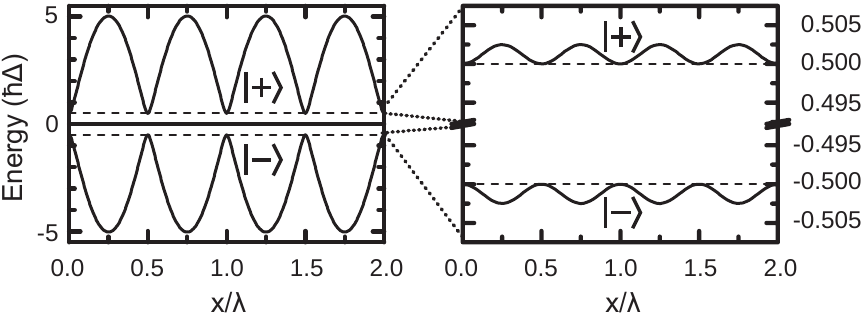}
\caption{{\it Left:} Dressed state potentials for $\omega_{R}/|\Delta| =10$.
Dashed lines indicate zero-intensity levels. Note the difference
in shape between the potential minima for $|+\rangle$ and
$|-\rangle$ states. {\it Right:} Same, for $\omega_{R}/|\Delta|
=0.1$. In this case both potentials are
equivalent.}\label{fig:pot}
\end{figure}

Non-adiabatic transitions can occur when the atom moves too fast, or in other
words, the eigenstates (dressed states) change too fast for the internal
state of the atom to follow. The atom is then transferred from one dressed
state to the other. This is most likely to occur when atoms cross a node of
the standing wave. Non-adiabatic transfer is incorporated in the simulation
procedure, but does not contribute significantly for present experimental
conditions.

Before we show the results of these semiclassical simulations, we discuss the
potential that is used. The left part of Fig.~\ref{fig:pot} displays the
potential for the case of a strong light field, where $\omega_{R}/|\Delta|
=10$. The case of a weak light field with $\omega_{R}/|\Delta| =0.1$ is
displayed to the right. Both potentials are very different in shape.

In the strong field limit, the potential minima of the $|+\rangle$ state
differ radically in shape from those of the $|-\rangle$ state. The minima of
the $|-\rangle$ state are smooth, broad and parabolic over a large range
around the minima. Using such a potential is a way to avoid non-parabolic
aberrations in the atom focusing process. On the other hand, the minima of
the $|+\rangle$ state look like a V-shaped potential. The oscillation period
of the atoms in the potential then depends on the starting point of the
oscillation, leading to strong aberration in the focussing. The optical
equivalent of such a lens is a conical lens or axicon. The aberration leads
to a less sharply defined focus with a long focal depth, or a focal line.

For positive detuning, the atoms start out in the $|+\rangle$ state and are
subject to the V-shaped potential. Provided that they do not switch between
dressed states due to spontaneous emission, this then creates focal lines --
aligned with the nodes -- rather than focal points. Hence, the formation of
structures should be fairly insensitive to the experimental parameters. For
negative detuning, the situation is reversed and the atoms are focussed by
the smooth potential, in principle creating sharper focal points -- aligned
with the antinodes -- because of the reduced non-parabolic aberration.
However, these focal points are subject to severe abberation caused by the
spread in longitudinal velocity, the atom-optical equivalent of chromatic
aberration. This negates the advantage of the reduced non-parabolic
aberration.

The potential for a weak standing wave is sine-like. The
difference in shape between the $|+\rangle$ and $|-\rangle$ state
minima has vanished, and focusing is expected to proceed in a
similar fashion for both states.

The experiments and simulations described here have all been performed in the
strong field case, with typical maximum values of $\omega_{R}/|\Delta|
\approx 5$, with both positive and negative detuning.

For the experiment, the $^{5}$D$_{4} \rightarrow ^{5}$F$_{5}$ atomic
transition of iron is used. The wavelength is 372~nm. The isotope $^{56}$Fe
has a natural abundance of 91.8\%, of which at a typical evaporation
temperature of 2000~K $\approx$50\% is still in the $^{5}$D$_{4}$ ground
state. Hyperfine structure is absent in $^{56}$Fe. For this transition,
$\Gamma=1.62\times 10^7\,{\rm s}^{-1}$ and the saturation intensity for the
$M_g=4 \rightarrow M_e=5$ magnetic subtransition is $I_s=65\,{\rm W/m^2}$.
The light mask is a one dimensional standing light wave produced by an
elliptical Gaussian beam with waist radius $w_x$ along the atomic beam axis
and $w_y$ perpendicular to it.

For the calculation, we start with 10,000 atoms. Each atom is initially
assumed to be in a random magnetic substate. The saturation intensity is then
calculated using the appropriate Clebsch-Gordan coefficient for the substate
and laser polarization. Spatially, the atoms are homogeneously distributed
over a single wavelength in the $x$-direction.

The longitudinal velocity distribution of the atomic beam is that of an
effusive beam, i.e., a Maxwell-Boltz\-mann distribution with average velocity
$\langle v \rangle=\sqrt{\frac{8k_BT}{\pi m}}$. The transverse velocity
distribution of the atoms is determined by the longitudinal velocity and the
geometrical collimation. The beam divergence, emerging from a round nozzle
with diameter $D$ at a distance $L$ from the standing wave, is characterized
in good approximation by a Gaussian angular distribution with a
root-mean-square (RMS) width of $\frac{D}{4L}$. This leads to a Gaussian
transverse velocity distribution with an RMS spread of $\frac{D}{4L}\langle v
\rangle$.

\begin{figure}[t]
\center
\includegraphics[width=85mm]{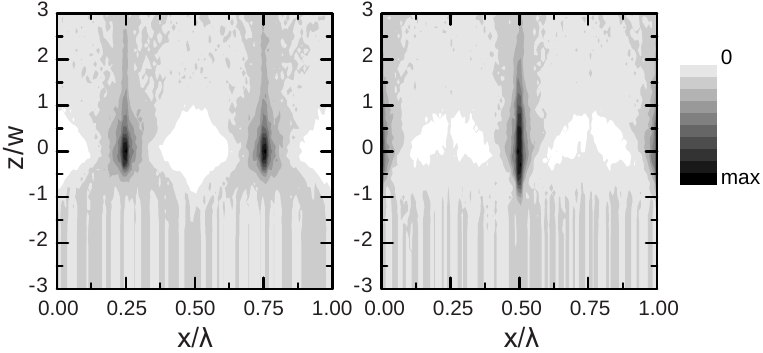}
\caption{Flux distribution of an atomic beam passing from below
through a strong light mask $\omega_{R}/|\Delta| =10$. {\it
Left:} $\Delta<0$ generates focal points, i.e., the size of the
focus is approximately one beam waist in the $z$-direction. {\it
Right:} $\Delta>0$ results in focal lines, i.e., the focus is
spread over more than two beam waists. This difference in flux
distribution is due to the difference in potential as shown on the
left side of Fig.\ \ref{fig:pot}.}\label{fig:focus}
\end{figure}

The equation of motion of every atom is integrated over a set time interval,
before we check for spontaneous and non-adiabatic transitions. The effect of
both transitions is to change from one dressed state to the other. The
possibility that atoms, which undergo a spontaneous emission, decay to a
different magnetic substate, is neglected.

The integration starts when the atom's longitudinal position ($z$) is three
times the waist radius of the Gaussian laser beam before the center of the
laser beam, and ends at the same distance after the center. We make a
histogram of the atomic flux distribution at set longitudinal positions.

To demonstrate the difference in focusing between the negative and positive
detuning, we first simulate focusing of a perfectly collimated atom beam,
setting the RMS spread in transverse velocity to zero. A round laser beam
with $w_y=w_z=50\,\mu{\rm m}$, a power of $50\,{\rm mW}$ and linear
polarization is taken. The detuning of the light with respect to the atomic
transition is set at $\Delta = \pm\,150$ (2$\pi$)~MHz ($\pm\,$58~$\Gamma$).
These parameters lead to a maximum value of $\omega_{R}/|\Delta| \approx 5.9$
The results are shown in Fig.~\ref{fig:focus}. Interaction with a red-detuned
standing wave generates focal points, as shown on the left. The blue-detuned
standing wave gives rise to focal lines, as shown on the right. For blue
detuning, the atoms experience on average 0.45 dressed state-changing
spontaneous emissions occur. For red detuning, the average number of
emissions is 0.7, reflecting the fact that atoms are attracted to intensity
maxima in this case, where the absorption rate is higher.

The strength of the lenses varies per atom. This is due primarily to the
distribution over the magnetic substates of the Fe atoms; secondarily, the
longitudinal velocity spread of the Fe atoms also contributes. The relatively
long $z$-range over which atom focusing occurs for the blue detuned case
stands out.

\section{Experimental setup}
\label{expsetup}

An Fe atomic beam is produced using an Al$_{2}$O$_{3}$ crucible with a nozzle
diameter of 1~mm heated to a temperature of around 2000~K by a carbon spiral
heater \cite{art16}, resulting in a typical Fe density in the source of
$n_{Fe}=4 \times 10^{20}$~m$^{-3}$. At this temperature the Fe-beam intensity
is $I_{Fe}=2.5 \times 10^{16}$~s$^{-1}$sr$^{-1}$, and the average
longitudinal velocity $\langle v \rangle = 870\,{\rm m/s}$.

The 372~nm light needed to excite the $^{5}$D$_{4} \rightarrow ^{5}$F$_{5}$
transition is produced by a titanium-sapphire laser, frequency doubled with
an LBO-crystal in a ring cavity. This laser system is locked to the
transition using a hollow cathode discharge cell, described in \cite{art17}.
An Acousto-Optical Modulator (AOM) is used to introduce the detuning of
$\Delta = \pm 150(2\pi$)~MHz between the light and the atomic resonance.

\begin{figure}[htbp]
\center
\includegraphics[width=85mm]{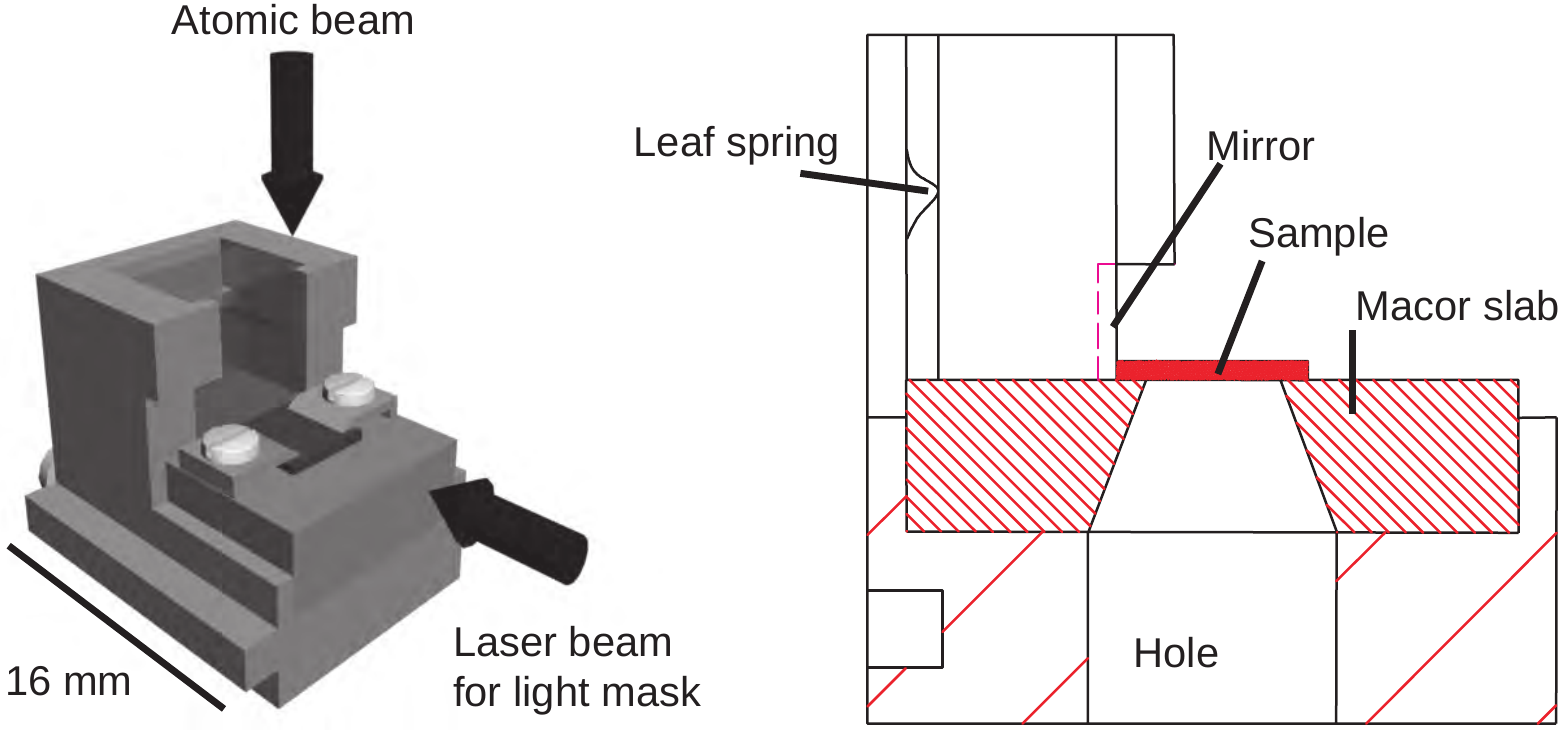} \caption{Left:
sample holder. Right: cross section of sample holder. The Si
sample is clamped to a ceramic isolator that is screwed to the steel
frame. The mirror is pressed to the frame by a spring.}
\label{sample}
\end{figure}

The Fe atoms are deposited on a native oxide Si[100] substrate, 650~mm
downstream from the nozzle. With the 1~mm nozzle hole diameter, the
divergence of the Fe beam therefore amounts to $\alpha_{RMS}=0.4\,{\rm mrad}$
RMS and the spread in transverse velocity to $0.35\,{\rm m/s}$ RMS. The
divergence is thus considerably larger than the best value obtained in our
earlier experiments with transverse laser cooling of Fe~\cite{art8}
($\alpha_{RMS}=0.17\,{\rm m/s}$). However, the divergence does not yet limit
the obtained width of the deposited nanolines, which is equal to the width
obtained with the lasercollimated beam~\cite{art5}.

The Si samples can be 3 to 8~mm wide and 2 to 10~mm long. The Si sample is
clamped on a compact sample holder (Fig.\ \ref{sample}), on which also an $8
\times 8 \times 3$~mm mirror is mounted. These sample holders can easily be
removed from the deposition chamber and stored in a load lock by a magnetic
translator so that the alignment of sample and mirror, which have to be
perpendicular to each other within 1~mrad, can be done \textit{ex vacuo}. The
pressure in the deposition chamber is typically $1 \times 10^{-8}$~mbar.

To create the standing light wave, the laser beam is focused to a waist of
50~$\mu$m parallel to the atomic beam and 80~$\mu$m perpendicular to the
atomic beam, located at the surface of the sample holder mirror. The light is
linearly polarized.

\begin{figure}[ht]
\center
\includegraphics[width=7cm]{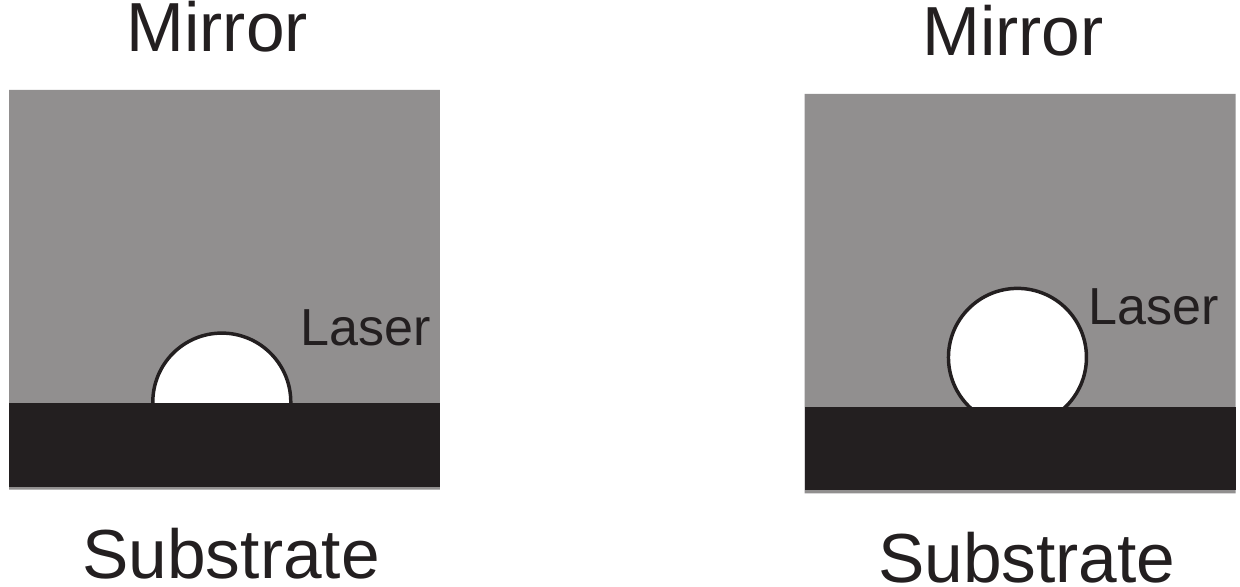} \caption{The
cut-off of the laser beam by the substrate. {\it Left}: The laser beam is cut
in half by the substrate (50\% cut-off). {\it Right}: Only 10\% of the laser
beam is cut off by the substrate (10\% cut-off).} \label{cutoff}
\end{figure}

The substrate is positioned close to the center of the laser beam.
Experiments have been performed with the substrate at the center of the
waist, thus cutting the laser beam in half, as well as with the substrate
positioned at the rear end of the laser beam waist, such that just 10\% of
the power in the laser beam is cut off by the substrate. Both cases are
schematically drawn in Fig.\ \ref{cutoff}.
According to the simulations, the 50\% cut-off configuration should lead to
somewhat narrower focussed lines and results that are less sensitive to
variations in laser power. However, diffraction effects, which are difficult
to take into account in the calculations, are more serious than in the 10\%
cut-off case.

To prevent Fe from oxidizing, a Ag capping layer of approximately 5~nm is
deposited using an effusive source operated at a typical temperature of
1140~K, resulting in a deposition rate of 0.15~nm/min.

\begin{figure}[ht]
\center
\includegraphics[width=85mm]{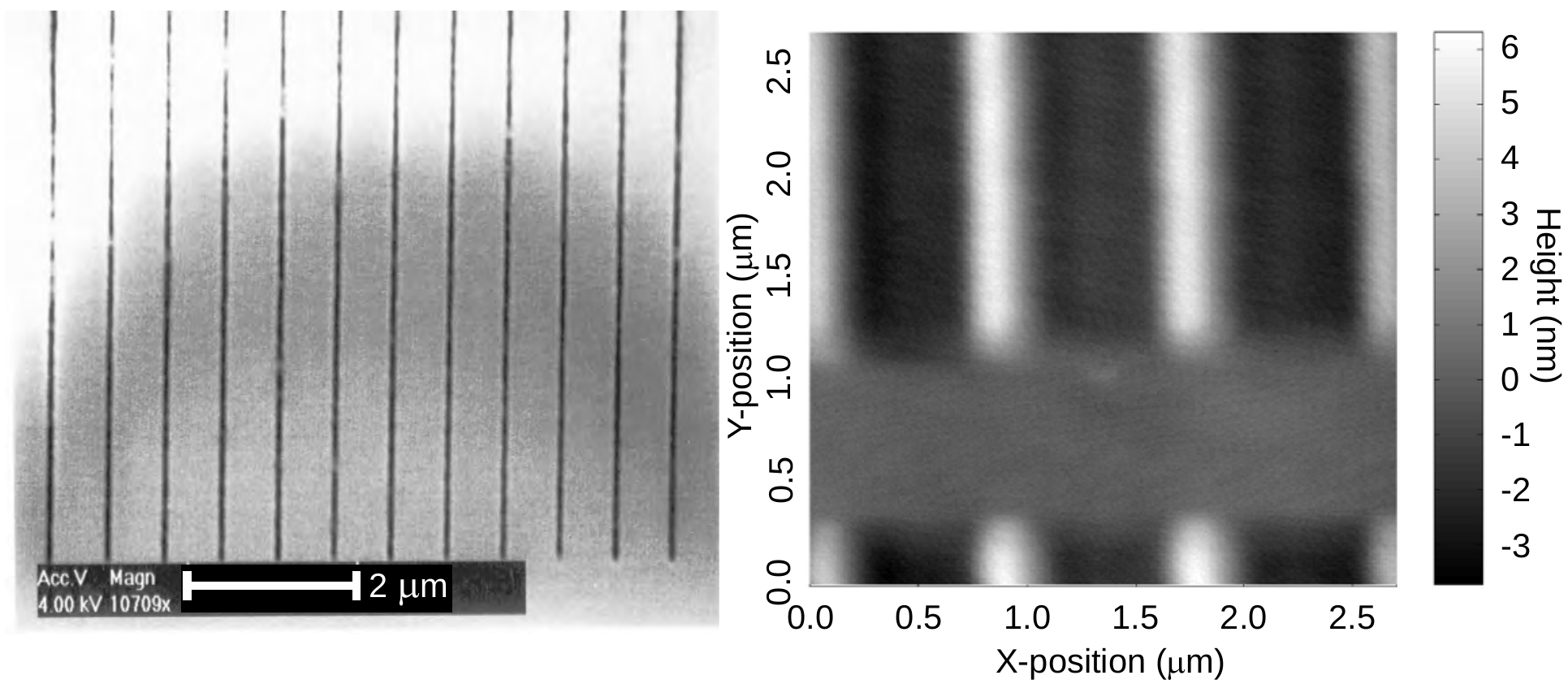} \caption{{\it
Left:} SEM-scan of the SiN mask. The pitch of the lines is 744~nm.
{\it Right:} AFM-scan of the deposited Fe-lines through the mask.
Deposition time is two hours.} \label{mask}
\end{figure}

Although the height of the deposited structures can be easily measured by an
AFM microscope, in order to determine the average thickness of the deposited
Fe layer as well (and hence the thickness of the background layer) we need to
calibrate the overall Fe deposition rate. To achieve this, we first produce
structures by depositing Fe through a mechanical mask. The mask is made by
e-beam lithography of a SiN membrane. Lines have been etched in the membrane
with a width of about 150~nm and a period of 744~nm over an area of 250
$\times$ 250~$\mu$m. The mask is pressed onto the substrate with a metal foil
of 100~$\mu$m as spacer between the bottom of the mask and the front of the
substrate. Deposition through such a mask results in background-free
nanolines, and thus gives a direct measurement of the total deposited layer
thickness. A SEM-scan of the mask and an AFM-scan of the Fe-lines grown
through the mask are shown in Fig.\ \ref{mask}. The FWHM width of the
structures is about 150~nm. From the height of the deposited structures the
deposition rate is measured. The measured deposition rate varies between 3
and 7 nm/hour, depending on the exact temperature and filling of the
crucible. In the evaluation of the deposited structures, the calibration of
the deposition rate has been performed under conditions as close to the
specific deposition experiment as possible.

\section{Results}
\label{exp}

We have deposited nanoline arrays of Fe with both negative and positive
detuning of the light and with substrates positioned both at 50\% cut-off and
at 10\% cut-off. The dependence on the light intensity of the height and
width of the nanolines has been investigated, as well as the ratio of line
height to the average layer thickness. Since we have an AOM with a fixed
RF-frequency of 150(2$\pi$)~MHz, our measurements are limited to blue and red
detuned focusing at this frequency shift from resonance.

\begin{figure}[htbp] \center
\includegraphics[width=70mm]{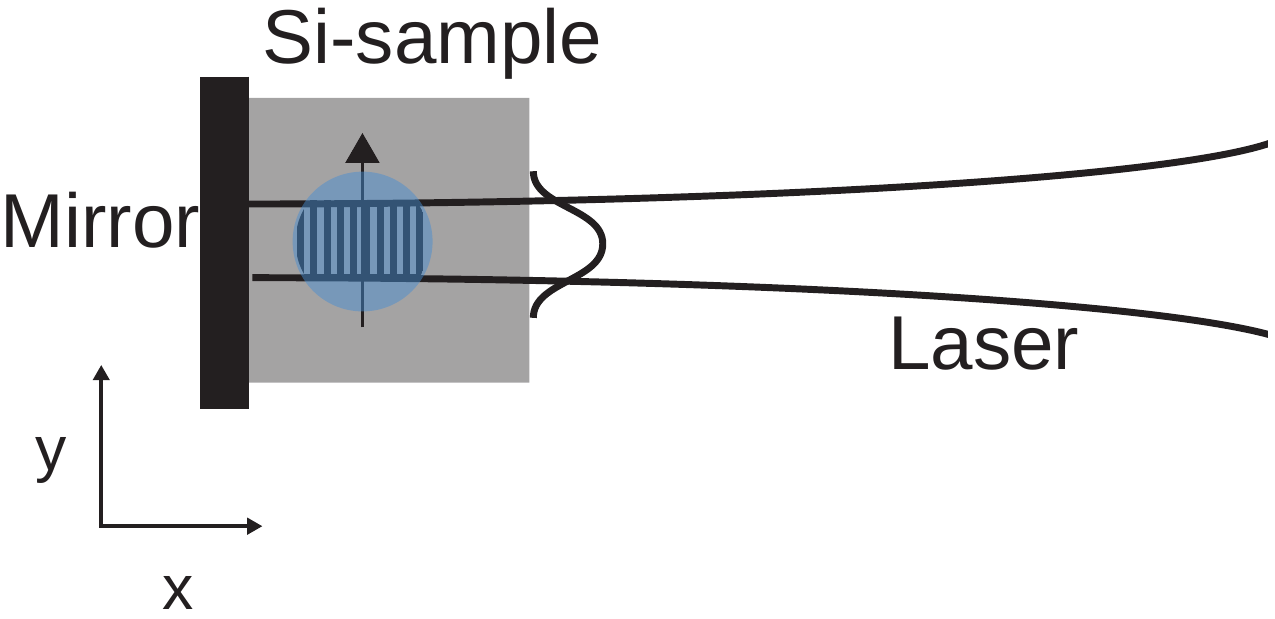} \caption{Deposition assembly
viewed from the direction of the incoming atomic beam. Nanolines are deposited on the substrate
where the atomic beam intersects the standing light field (round spot). The
height and width of the nanolines are investigated as a function
of position along the nanolines. Since the laser beam is Gaussian,
each position corresponds to a specific light intensity.}
\label{gausslas}
\end{figure}

On each deposited sample, the height and width of the nanolines has been
measured as a function of position transverse to the laser beam axis (i.e.,
along the $y$-axis). Fig.~\ref{gausslas} depicts the geometry. As the laser
beam has a Gaussian profile, each $y$-position is characterized by a
different light intensity.

\begin{figure}[htbp]
\center
\includegraphics[width=70mm]{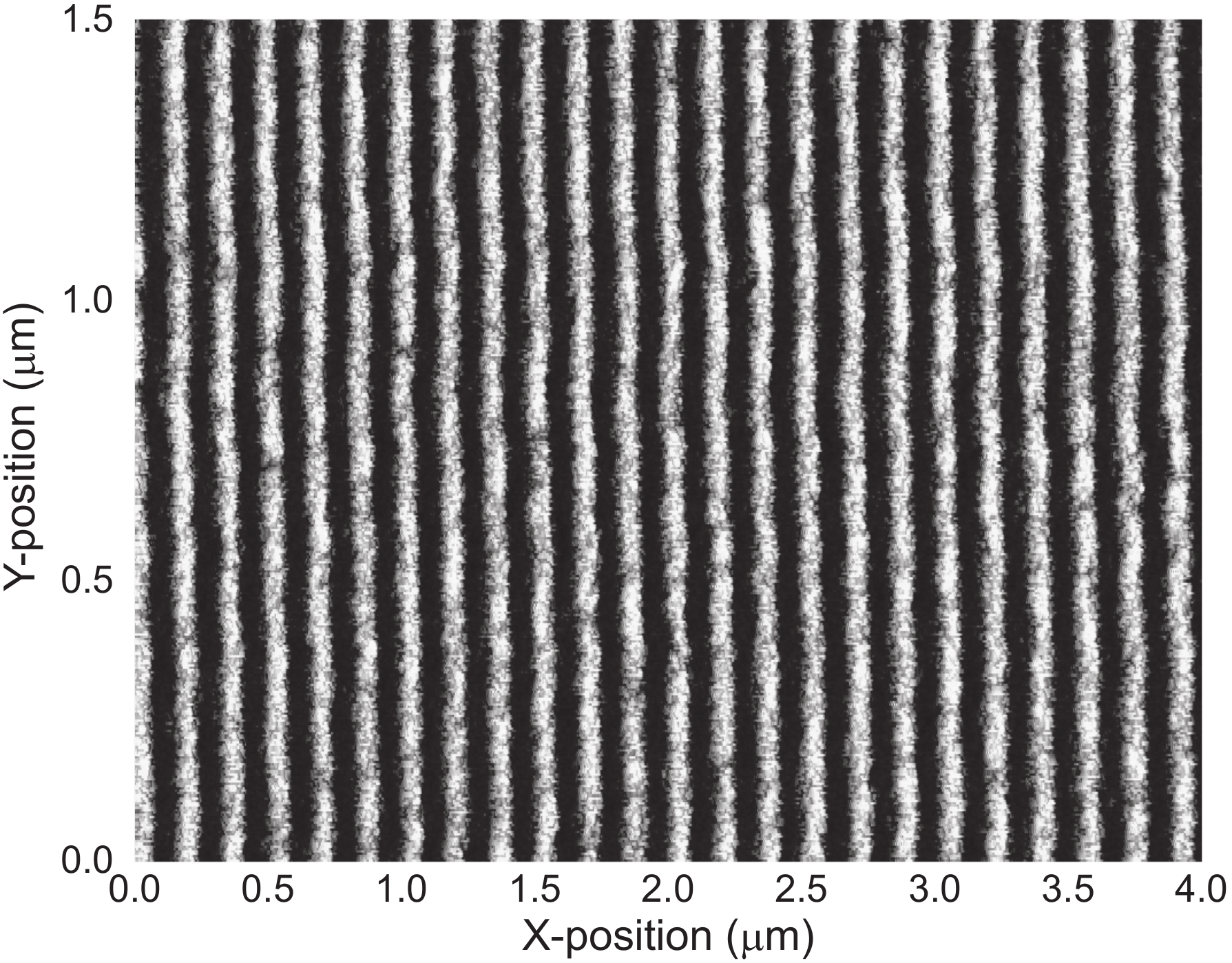}
\caption{Example of a 4 $\times 1.5$~$\mu$m AFM-scan of the nanolines. The height (grey)scale ranges over 6~nm.}
\label{AFM}
\end{figure}

\begin{figure}[htbp]
\center
\includegraphics[width=70mm]{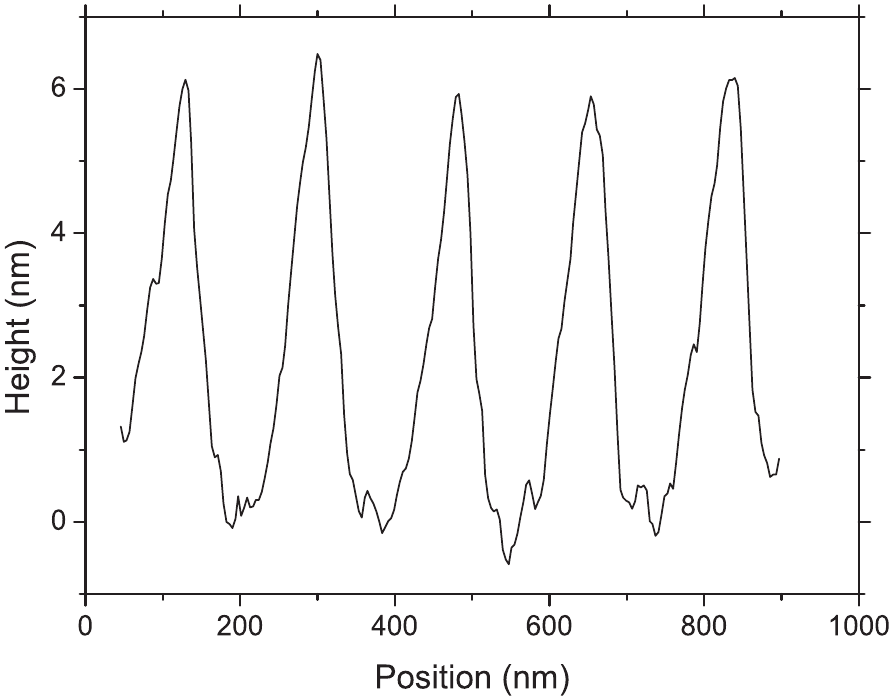}
\caption{1~$\mu$m line-scan, averaged over 40~nm in the
Y-direction, of the same sample shown in Fig.\ \protect{\ref{AFM}}. The deposition time is 2 hours. The height of the
structures in this image is 6~nm and their FWHM width
approximately 60~nm. The laser power is $P$=50~mW, 10\% of the
laser beam is cut-off by the Si substrate.}
\label{cros}
\end{figure}

In Fig.\ \ref{AFM} a 4 $\times 1.5$~$\mu$m AFM scan of a typical sample after
deposition is shown. In 2 hours deposition time we grow structures up to
heights of 6~nm. Fig.\ \ref{cros} shows a cross section through the scan over
a range of $1\,\mu{\rm m}$. The distance between consecutive lines (the pitch
of the modulation) is equal to $\lambda/2$=186~nm.

\begin{figure}[t]
\center
\includegraphics[width=7cm]{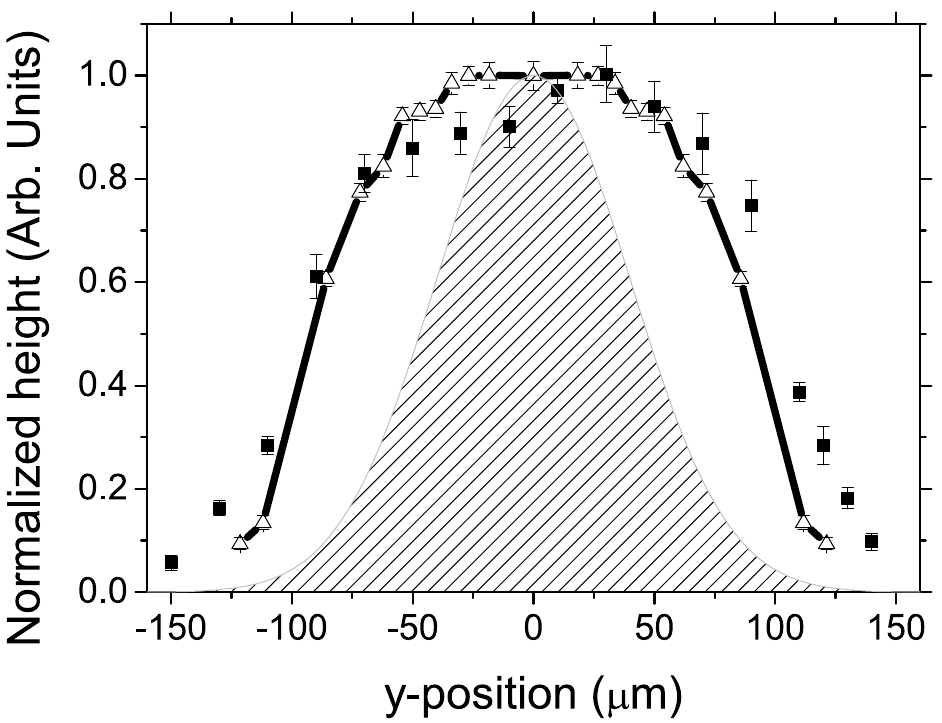}
\caption{Typical height profile as a function of position along the
lines for a blue detuned 50\% cut-off laser beam. Both simulation
data and experimental data are normalized to their maximal height.
The squares are the measurements, the triangles are the
simulation. The patterned profile reflects the dependence of the
standing wave intensity on position. The laser power is
$P$=50~mW.} \label{height}
\end{figure}

Fig.\ \ref{height} shows a typical profile of the height of the nanolines as
a function of transverse position. The intensity profile of the laser beam is
shown as well. For high intensities, the height of the nanolines is largely
independent on the light intensity both in experiment and simulation: the
height profile does not follow the Gaussian shaped laser intensity profile,
but it has a broad flat top. The height of the potential shown in Fig.~
\ref{fig:pot} follows, in the high intensity limit, from the square-root of
the intensity. Increasing the intensity will move the focus over the $z=0$
point towards negative $z$-values. Since the lens, especially in the blue
detuned case, has a large focal depth, the atoms remain focused at the $z=0$
position, even for higher intensities, resulting in the flat top of the
height profile.

The results of the simulations are shown as well in Fig.~\ref{height}. Both
curves are normalized to the same height. The agreement between the
experimental data and the results of the simulations is good when we look at
the shape of the profile. At low intensities, the measurements deviate
somewhat from the simulations. This may be caused by the non-Gaussian tails
in the spatial profile of the laser beam used in the experiment.

\begin{figure}[ht]
\center
\includegraphics[width=85mm]{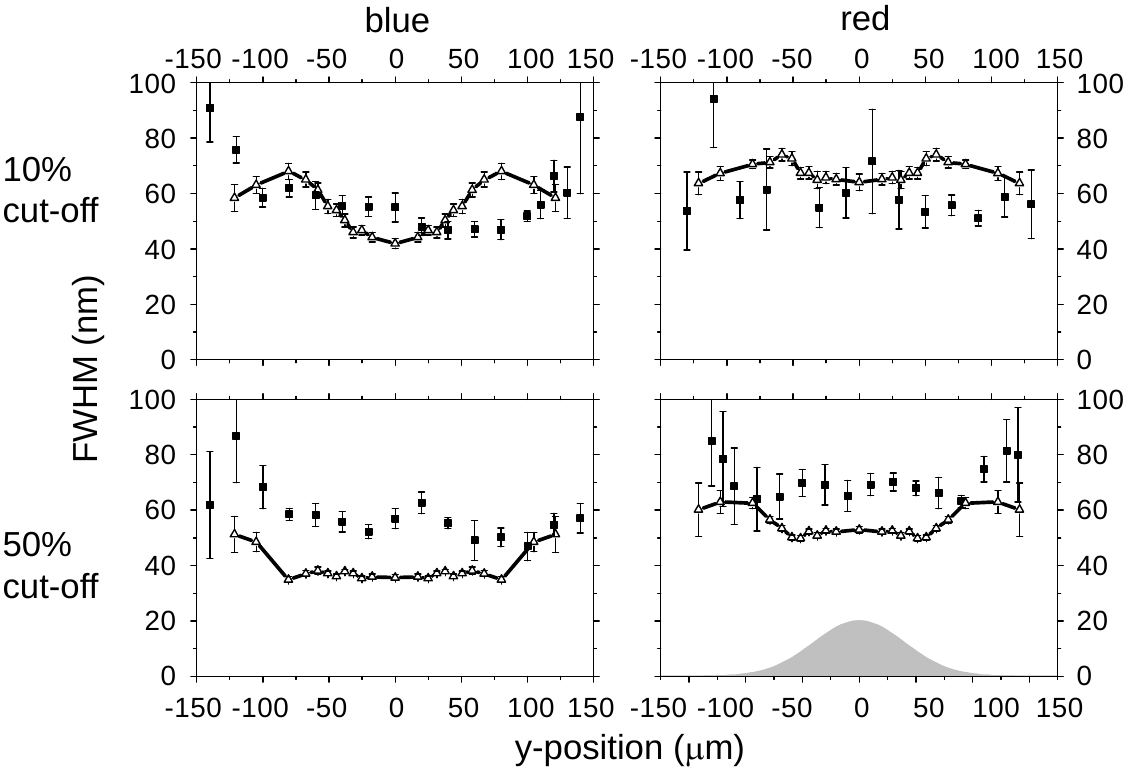} \caption{FWHM   of
the lines as a function of position along the lines for blue and
red detuning. The squares are the measurements. Simulations are
indicated by the triangles. In the lower right graph, the laser
intensity profile as used in the simulations is shown.}
\label{width}
\end{figure}

Experimental data and simulation results on the width of the structures as
function of the position on the substrate are shown in Fig.\ \ref{width}.
For high intensities, blue detuning results in slightly narrower lines
compared to red detuning in the experiment. This difference is more
pronounced in the simulations, and results from different focusing potentials
(Fig.\ \ref{fig:pot}). The match between measured and simulated widths is
better in the 10\% cut-off configuration than in the 50\% cut-off
configuration. Diffraction of the laser beam is not included in the
simulation. This may result in an underestimate of the nanoline widths in the
50\% cut-off simulation.

Determining the ratio of the height of the nanolines to the average thickness
of the Fe layer requires the calibration of the overall deposition rate as
discussed in Section \ref{expsetup}. Although this calibration is not very
accurate, the result is that in our experiment the ratio for all experimental
conditions is approximately 0.5 for the lines at maximum laser intensity (the
center of the height profile as shown in Fig.~\ref{height}).

\begin{table}[h]
\begin{center}
\caption{Simulation results of the ratio of the height of the
lines to the average layer thickness.} \label{tabcont}
\begin{tabular}{lcc}
\hline \hline
&red&blue\\ \hline
50\% cut-off&1.25&1.25\\
10\% cut-off&0.60&0.70\\ \hline
\end{tabular}
\end{center}
\end{table}

To obtain the same ratio from the simulations we assume that 50\% of the
atoms are in the $^{5}$D$_{4}$ ground state, which is the thermal occupation
of the ground state at a source temperature of 2000~K. The other 50\% of the
$^{56}$Fe atoms, as well as the other isotopes, are not affected by the
focusing light field and only contribute to the background layer. In Table
\ref{tabcont} the ratio between simulated line heights and the average Fe
layer thickness at maximum laser intensity is listed. For the 10\% cut-off
setting the simulated ratio is close to the experimental value. For the 50\%
cut-off case, the experimental ratio is much lower than the simulated ratio.
This discrepancy between measurement and simulation for the 50\% cut-off case
is also present in the widths of the lines: The width of the simulated
nanolines is smaller than the width of the deposited nanolines. This means
that the quality of the deposited nanolines in the 50\% cut-off case is lower
than what can ideally be achieved. This may be due to the diffraction of the
laser beam in the 50\% cut-off case, which is not included in the
simulations.

An analysis of the magnetic properties of the deposited nanolines will be
published elsewhere~\cite{Fabrie2009}.

\section{Conclusions}
\label{concl} By direct-write atom lithography Fe-nanolines are deposited
with a pitch of 186~nm and a FWHM width of 50~nm, without the use of laser
collimation techniques. Sufficient collimation is obtained by strong
geometrical collimation, produced by the relatively large distance (650~mm)
between Fe-source and sample compared to the 1~mm source nozzle. In this way
the divergence of the atoms arriving at the sample is reduced to 0.4~mrad
RMS.

Experiments and simulations are performed with the focusing standing light
wave cut in half by the substrate and with a 10\% cut-off by the substrate.
Experiments and simulations are in good agreement with respect to the height
of the nanolines. The experimental width of the nanolines corresponds well
with the simulations for the 10\% cut-off case. For the 50\% cut-off case the
measured width of the lines is slightly larger than the simulated value.
Also, the measured height of the lines versus average deposited layer
thickness ratio is lower than in the simulations in that case. We attribute
this discrepancy to diffraction of the light beam by the substrate and small
misalignments of the standing light wave.

The fact that direct-write atom lithography can be applied without the use of
laser cooling techniques opens the way to many new applications. Except for a
small number of elements with a strong closed-level transition in an easily
accessible wavelength region, laser cooling can add a serious and sometimes
insurmountable complication to atom lithography. Past efforts to apply the
technique to atoms that are interesting from a technological viewpoint (e.g.,
gallium or indium in view of III-V semiconductor applications) have
concentrated on the laser collimation as a first step. These efforts have
succeeded in the challenging goal of achieving laser
collimation~\cite{art9,Kloeter2008}, but have not yet resulted in the
production of nanostructures. The presented results show that, by using an
high-flux beam source and relying on geometrical collimation, laser
collimation can be bypassed completely.

\begin{acknowledgement}
This work is part of the research program of the Foundation for Fundamental
Research on Matter (FOM), which is financially supported by the Netherlands
Organisation for Scientific Research (NWO).
\end{acknowledgement}

\end{document}